\documentclass[pra,twocolumn,preprintnumbers,floatfix]{revtex4}
\usepackage{graphicx}
\usepackage{dcolumn}
\usepackage{bm}
\usepackage{latexsym,epsfig}
\usepackage{graphicx}
\usepackage{verbatim}
\usepackage{comment}
\usepackage{amsmath}
\usepackage{amssymb}
\usepackage{stmaryrd}
\usepackage{color}
\usepackage{epstopdf}
\usepackage{grffile}
\usepackage{ulem}
\DeclareGraphicsExtensions{.eps}
\newcommand{\beq}{\begin{equation}}
\newcommand{\eeq}{\end{equation}}
\newcommand{\bea}{\begin{eqnarray}}
\newcommand{\eea}{\end{eqnarray}}
\newcommand{\ben}{\begin{eqnarray*}}
\newcommand{\een}{\end{eqnarray*}}
\newcommand{\bfig}{\begin{figure}}
\newcommand{\efig}{\end{figure}}

\usepackage{hyperref}
\hypersetup{
    colorlinks=true,      
    urlcolor=blue,
    citecolor=blue,
    linkcolor=blue
}

\newcommand{\tapan}[1]{\textcolor{blue}{\bf #1}}

\definecolor{augustine}{rgb}{1,0,1}

\begin{document}
\title{Two-body repulsive bound pairs in multi-body interacting Bose-Hubbard model}

\author{Suman Mondal$^{1}$, Augustine Kshetrimayum$^2$ and Tapan Mishra$^{1}$}
\affiliation{$^{1}$Department of Physics, Indian Institute of Technology, Guwahati, Assam - 781039, India}
\affiliation{$^2$ Dahlem Center for Complex Quantum Systems, Physics Department, Freie Universit\"{a}t Berlin, 14195 Berlin, Germany}
\date{\today}

\begin{abstract}
We study the system of multi-body interacting bosons on a two dimensional optical lattice and analyze the formation of bound bosonic pairs in the 
context of the Bose-Hubbard model. Assuming a repulsive two-body interaction we obtain the signatures of pair formation in the regions 
between the Mott insulator lobes of the phase diagram for different choices of higher order local interactions. 
Considering the most general Bose-Hubbard model involving local multi-body interactions we investigate the ground state properties utilizing 
the cluster mean-field theory approach and further confirm the results by means of sophisticated infinite Projected Entangled Pair States calculations. 
{By using various order parameters, 
we show that the choice of higher-order interaction can lead to pair superfluid phase in the system between two different Mott lobes.} 
We also analyze the effect of temperature and density-dependent tunneling to establish the stability of the PSF phase. 
%

\end{abstract}


\maketitle
\section{Introduction}
The seminal observation of quantum phase transition between the superfluid(SF) and the Mott insulator(MI) phases 
in optical lattice~\cite{bloch} and its theoretical prediction~\cite{Fisher89,jaksch} 
have revolutionized the field of strongly correlated quantum matter. The physics which emerges out of the competition between 
the local two-body interactions and the 
off-site hopping strengths of the paradigmatic Bose-Hubbard(BH) model is regarded as one of the simplest examples of quantum simulations. 
The underlying mechanism which drives this interesting phase transition is the high-level 
of tunability of on-site interactions with respect to the hopping amplitudes using the technique of the 
Feshbach resonance and/or the lattice strengths. Following this 
experimental observation, several interesting phenomena have been unveiled at the interface of atomic, molecular, optical and condensed matter physics 
in recent years considering 
many variants of the BH model. However, the simple BH model with only on-site interactions itself have revealed a plethora of exotic 
physics in different contexts ~\cite{BlochRev,kennett2013,LewensteinBook}. 

{Recently, effective higher-order interactions have been observed in optical 
lattice experiments ~\cite{wills,Mark2011}. These effective interactions are due to 
the virtual population of occupation dependent higher Bloch bands. Although these 
effects are small compared to the original two-body interactions, they provide 
enough motivation to explore the physics of ultracold matter in the presence of 
multi-body interactions in optical lattices.} An immediate usefulness of such multi-body interactions 
can be understood in the context of the attractive BH model which involves local 
two-body attractive interactions. It has been shown that for any finite attractive interaction the bosons occupy 
a single site of an optical lattice leading to collapse~\cite{Dalfovo99}. 
This difficulty can be overcome by including a very strong three-body on-site repulsion which prevents the occupation of a lattice site by more 
than two atoms and hence the collapse. 
A recent proposal rigorously shows that an infinitely strong three-body repulsion 
can arise due to the three-body loss process resulting from the elastic scattering of 
atoms~\cite{daley}. This infinite three-body repulsion which is termed as the three-body hardcore constraint, facilitates the 
formation of attractively bound bosonic pairs. The superfluid of these composite pairs is called the pair superfluid(PSF) phase in 
optical lattice~\cite{daley,wessel,singh1,Yung2011} 
which is an interesting manifestation of competing two and three-body interactions. Several other 
theoretical proposals have been made recently to control the three-body interactions in various ways in optical lattices~\cite{tiesinga,daley2,Sansone2012}. 
Moreover, recent proposal by D. Petrov~\cite{Petrov2} reveals the possibilities to 
simultaneously manipulate the higher order multi-body interactions along with 
the two-body one in atomic systems~\cite{Petrov1,Petrov2}. 
This prediction is one step forward in the directions of exploring physics arising due to the on-site interactions in optical lattices. 
With these types of interactions, 
the standard BH model gets modified accordingly and one gets
a more general BH model with the on-site multi-body interactions given as;  
\begin{eqnarray}
H=&-&t\sum_{<i,j>}(a_{i}^{\dagger} a_{j}+H.c.)\nonumber\\ 
  &+&\sum_{p=2}^M \Big({U_p}\sum_{i}{({n_i})!\over{p!(n_i-p)!}} \Big) -\mu \sum_{i} n_i\nonumber\\
\label{eq:ham1}
\end{eqnarray}  
where ${a_i}^{\dagger}({a_i})$ is the bosonic creation (annihilation) operator, $n_i$ is the number operator for the $i^{th}$ site,
and $<i,~j>$ denotes the nearest neighbor sites. While $t$ represents the nearest-neighbor hopping amplitude,
$U_p$ is the on-site $p$-body interaction strength. Depending on the value of $M$, one gets the corresponding multi-body interacting BH model. 
$\mu$ is the chemical potential associated to the system in the grand canonical ensemble which decides the number of particles in the system. 
As mentioned before, this model with only two body interaction $U_2$ exhibits the 
SF-MI phase transition at integer densities. As a result one gets the well known MI lobes corresponding to different atom densities 
in the ground state phase diagram plotted in the $\mu$ and $U_2$ plane. Hereafter, we denote the MI lobes for different particle densities 
as MI($\rho$) where $\rho$ is the ratio between the total number of particles to total number of sites in the systems.

Although, competing multi-body interactions in the BH model may provide interesting physics, the system with up to the three-body interactions($U_3$) has 
been widely studied in recent years~\cite{daley,wessel,MSingh1,YZhang,TSowinski1,sowinski1,Sansone2012,Ejima,Silva2012,Silva2014,Hincapie2016} revealing various 
interesting physical phenomena in optical lattices. However, in an interesting proposal in Ref.~\cite{Sansone2012}, it is shown that the strength of the three-body interaction 
$U_3$ can be tuned by coupling it to the Efimov states which leads to a non-trivial form of the interaction {$U_3\delta_{n,3}$}. 
The phase digram of the BH model in presence of such three-body interaction is obtained by using the simple mean-field theory approach analysis and 
complemented by the  Quantum Monte Carlo(QMC) calculation. This reveals that for attractive $U_3$ and repulsive $U_2$ the system favors a direct first order transition from the 
MI(1) to the MI(3) phase by completely suppressing the MI(1) lobe when {$|U_3/U_2| > 1$} even at finite temperature~\cite{Sansone2012}. However, this finding was later found to be 
inconsistent when compared to the density matrix renormalization group(DMRG) and the cluster mean field theory(CMFT) calculations in one and two dimensional 
systems respectively by some of us in Ref.~\cite{Singhpsf}. A careful analysis in Ref.~\cite{Singhpsf} showed that there 
exists no first order transition between the Mott lobes for the parameter choice considered in Ref.~\cite{Sansone2012}. 
Rather, the competing two and three body interactions lead to the formation of a non-trivial PSF phase in between the MI(1) and MI(3) lobes 
where the bosons tend to move in pairs even in the presence of the two-body repulsive interactions. This reveals a
 kind of two-body repulsively bound pairs driven by a mechanism completely different from the one observed 
in optical lattices by Winkler {\textit et al.}~\cite{Winkler2006} where the pair formation occurs due to the 
competition between the two-body interaction $U_2$ and the bandwidth. 

In this paper we show that to achieve this anomalous pairing of bosons with two-body repulsion, 
it is not always necessary to consider the specific form of the three-body interaction as discussed in 
Ref.~\cite{Sansone2012, Singhpsf}. The most general BH model given in Eq.~\ref{eq:ham1} with suitable choice 
of multi-body interactions may stabilize the 
PSF phase between the Mott lobes which will be discussed in more detail below. 
The remaining part of the paper is organized as follows. In Sec-II we explain the model considered for this work with a brief 
information about the methods. In Sec-III we discuss our results in detail and 
finally we conclude in Sec-IV.

\section{Method}
We numerically investigate the model shown in Eq.~\ref{eq:ham1} by restricting up to four-body interactions for simplicity. The 
explicit form of the Hamiltonian with all the interactions is given as;
\begin{eqnarray}
H=&-&t\sum_{\langle i,j\rangle}(a_{i}^{\dagger} a_{j}+H.c.)+{{U_2}\over{2}}\sum_{i}n_i(n_i-1)\nonumber\\
&+&\frac{U_3}{6}\sum_i n_i(n_i-1)(n_i-2)\nonumber\\
&+&\frac{U_4}{24}\sum_i n_i(n_i-1)(n_i-2)(n_i-3)\nonumber\\
&-&\mu\sum_{i} n_i
\label{eq:ham2}
\end{eqnarray} 
where the terms have their usual meaning as discussed before. 
In order to analyze the ground state properties of Eq.~\ref{eq:ham2} we 
first utilize the self-consistent CMFT approach which is an approximation method based on the simple 
single site mean-field theory approach~\cite{McIntosh,danshita,hassan}. In this case, the Hamiltonian is divided into several clusters of 
finite number of sites and each cluster interacts with the rest of the system in a mean-field way i.e.
\begin{eqnarray}
H_{CMF}=\sum_{i,j\in C}H_C ~-~ t\sum_{\substack{\langle i,j\rangle \\ i\in B, j\notin C}} (a_i^\dagger \psi_j + H.c.)
\label{eq:ham3cmft}
\end{eqnarray}
Here, $H_C$ is the cluster Hamiltonian identical to Eq.~\ref{eq:ham2} with index $i,~j$ belonging to the cluster $C$. The second term which is the mean-field expression for the hopping term 
from the $i^{th}$ site at the cluster boundaries($B$) to the nearest neighbor~\cite{Tomadin2010}. $\psi$ is the superfluid order parameter which is determined self consistently. 
In order to obtain the insights about various quantum phases we utilize the average density and the 
superfluid density of the system $\rho=1/L\sum_i n_i$ and $\rho_s=1/L\sum_i \psi_i^2$ respectively computed from the CMFT ground state where $L$ is number of sites in a cluster. 
It is well known that the CMFT method is more accurate than the simple mean-field field theory approach and can capture the 
qualitative picture of the system with less computing effort 
than the powerful QMC method~\cite{McIntosh,danshita,hassan,Singhpsf,singh1,luhmann}. Note that the accuracy of the method relies on the 
cluster size. In this case we consider a four site cluster which is sufficient to predict the relevant physics . 

In addition to the CMFT approach, we have employed the infinite Projected Entangled Pair States (iPEPS) algorithm which are two dimensional tensor network 
techniques \cite{iPEPSOld,2004peps}. Such techniques are built upon genuine quantum correlations and hence, goes beyond mean field calculations. 
Besides, we can directly target the thermodynamic limit by assuming translational invariance over some sites. Another advantage is that unlike QMC techniques, it does not suffer from the infamous sign problem for fermionic and frustrated systems~\cite{fermionicpeps}.
For these reasons PEPS techniques have been used in the past to study hard condensed matter problems such as 
frustrated kagome antiferromagnets\cite{Xiangkhaf,Thibautnematic,ThibautspinS,KshetrimayumkagoXXZ} and 
real materials \cite{Kshetrimayummaterial,CorbozMaterial,SSlandpeps4}. It has been able to beat state-of-the-art 
QMC calculations in finding the ground state energy of the doped Hubbard model~\cite{Corboz2DHubbard},  
and helped settle controversies that would have otherwise been difficult such as the magnetization plateaus of the 
Shastry-Sutherland model\cite{CorboztJ}, phase diagrams of steady states of dissipative spin models\cite{Kshetrimayum}, etc. 
The technique has now been extended to finite temperature calculations \cite{piotr2012,piotr2015,piotr2016,KshetrimayumThermal} and 
the difficult problem of time evolution in 2D~\cite{Claudiusevolution,piotrevolution,Kshetrimayumevolution,KshetrimayumTC}.
	
For the purpose of this work, we use an iPEPS with a two-site unit cell in the thermodynamic limit. 
We approximate the ground state of the Hamiltonian given in Eq.~\ref{eq:ham2} using the 
so-called simple update~\cite{simpleupdatejiang} with bond dimension $D=2$ and $D=4$ which proves sufficient for our purpose. {In the Results section, we provide the plots for $D=4$.}  
After analyzing the zero temperature phase diagram of the system we investigate the effect of thermal fluctuation in the system. The
finite temperature calculations are done with an annealing algorithm~\cite{KshetrimayumThermal} 
with infinite Projected Entangled Pair Operators (iPEPO) which are mixed state version of iPEPS~\cite{Kshetrimayum,KshetrimayumRMP}. 
The expectation values are computed using the Corner Transfer Matrix Renormalization (CTMRG) method~\cite{ctmroman2009,ctmroman2012}.


\section{Results and Discussion}
In this section we move on to discuss our results in detail which are obtained by using the CMFT and iPEPS approach for $M=4$ of Eq.~\ref{eq:ham1} . 
\begin{figure}[t]
\begin{center}
\includegraphics[width=1\columnwidth]{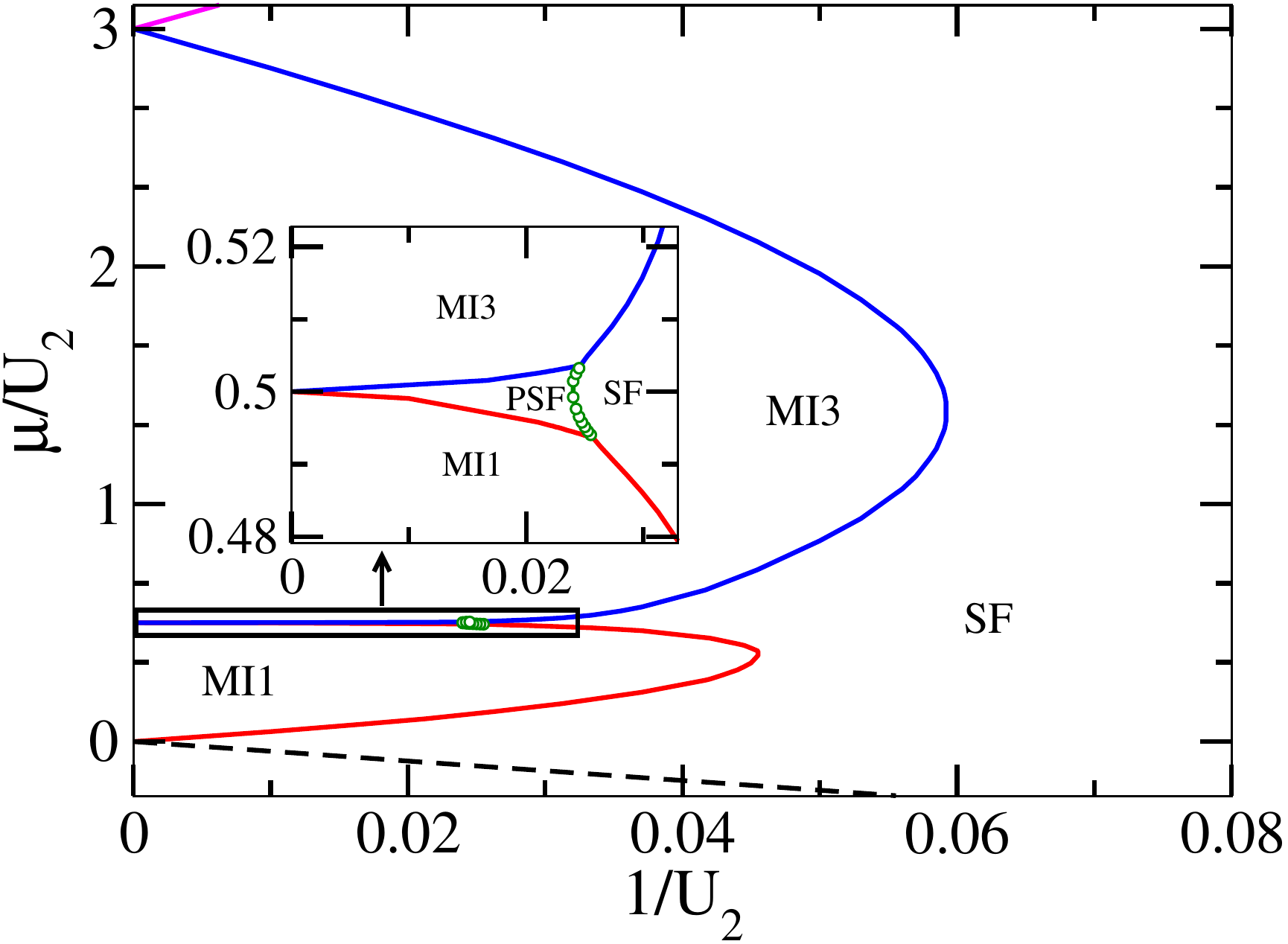}
 \end{center}
\caption{(color online)CMFT phase diagram for $t=1$, $U_3/U_2 = -2.0$ and $U_4/U_3 = -3.0$. Solid lines 
show the boundaries of MI phases and dashed line separates the empty state. Inset shows the PSF-SF boundary marked by green line with circles.}
\label{fig:phasedia1}
\end{figure}
\begin{figure}[b]
\begin{center}
\includegraphics[width=1\columnwidth]{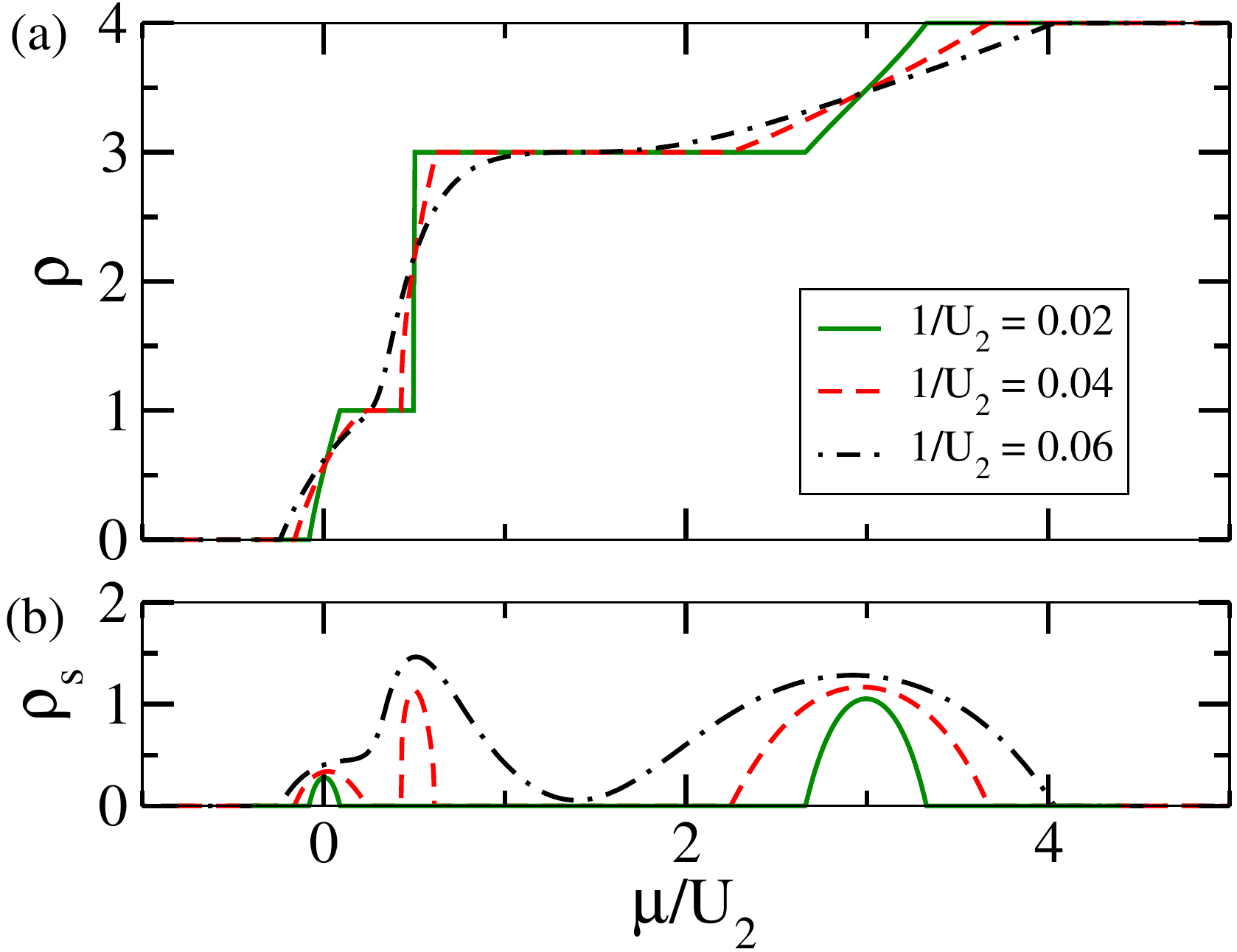}
 \end{center}
\caption{(color online)(a)$\rho$ vs. $\mu/U_2$ and (b)$\rho_s$ vs. $\mu/U_2$ plots for several cuts through the phase diagram of Fig.\ref{fig:phasedia1} corresponding to $0.02$(green solid line), 
$0.04$(red dashed) and $1/U_2=0.06$(black dot-dashed).}
\label{fig:rhomu1}
\end{figure}
\begin{figure}[b]
\begin{center}
\includegraphics[width=1\columnwidth]{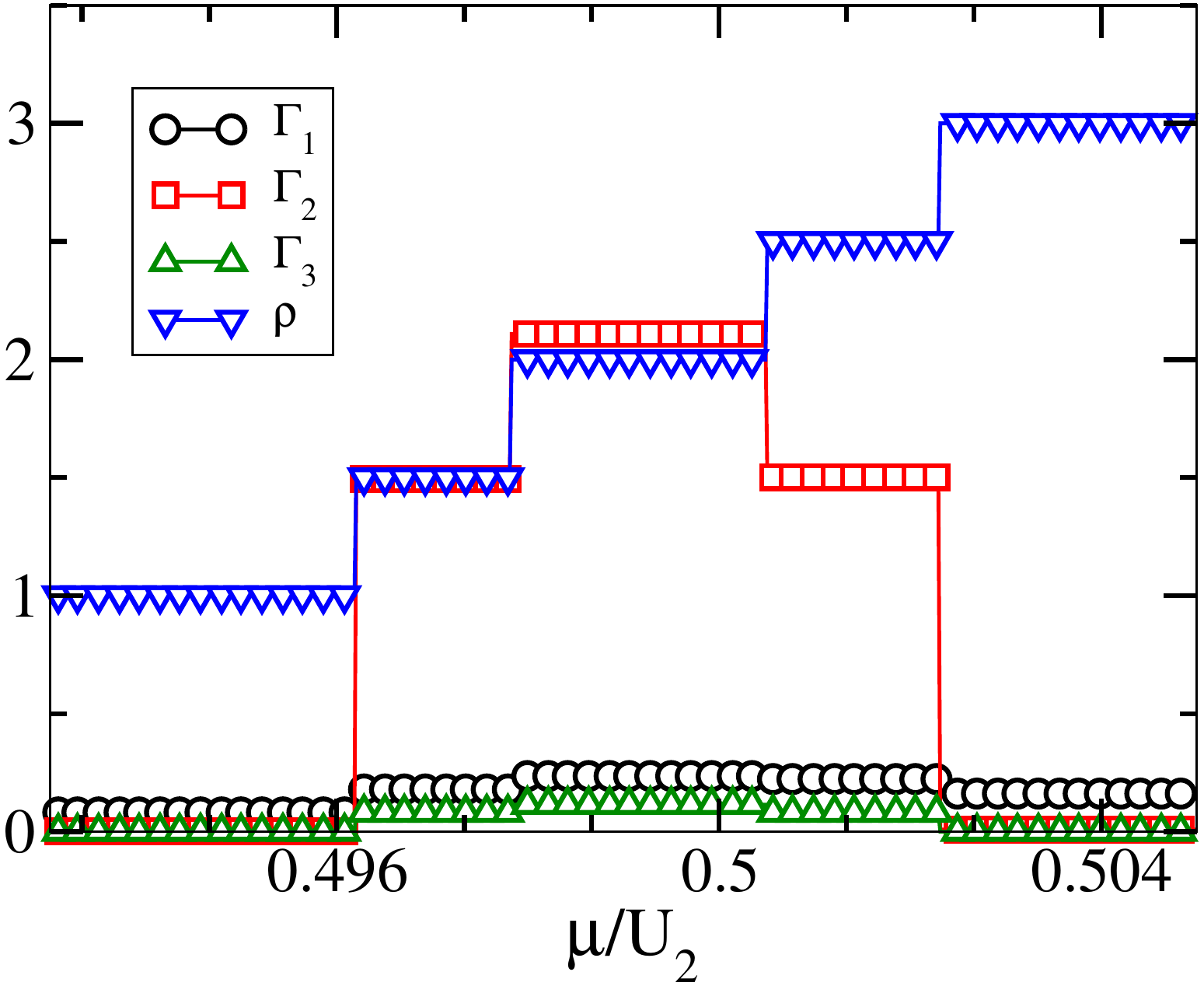}
 \end{center}
\caption{(color online)Correlation functions $\Gamma_n(i,j)$ for a cut  
along $1/U_2 = 0.02$ in the phase diagram Fig.\ref{fig:phasedia1}. The corresponding values of $\rho$(blue down triangles) are 
shown for comparison.}
\label{fig:corr1}
\end{figure}
Keeping terms up to $M=4$ in the model(\ref{eq:ham1}) we have three different interactions in the system such as $U_2,~U_3$ and $U_4$. Note that in our 
analysis the focus is to analyze the two-body repulsive bound pairs. Therefore, our obvious choice is to keep $U_2>0$. 
In this case, we consider attractive(repulsive) three-(four-)body interactions i.e. $U_3<0$ and $U_4 > 0$.
For simplicity, we define two ratios such as $U_4/U_3$ and $U_3/U_2$  and analyze the ground state phase diagram of the system. In the case of the BH model shown in Eq.~\ref{eq:ham1},
it is well known that the 
presence of interaction up to $U_3$ largely affects the SF-MI phase transitions with modified Mott lobes at higher densities. 
While the MI lobes corresponding to $\rho \geq  2$ get enlarged by the three-body repulsion $U_3$~\cite{TSowinski1,MSingh1}, 
an attractive $U_3$ results in shrinking up of the higher MI lobes~\cite{TSowinski1}. However, in this case we show that 
a large four-body repulsive interaction $U_4$ leads to interesting phenomena. {Note that the large $U_4$ is necessary to prevent the collapse due to attractive $U_3$ and also to stabilize the MI(3) state in the system}. In this case, the MI(3) lobe 
expands by simultaneously shrinking the MI(2) lobe which eventually disappears for some specific ratio of interactions defined above. 
In Fig.~\ref{fig:phasedia1}, we depict the phase diagram corresponding 
to the ground state of {Eq.~\ref{eq:ham3cmft} using the CMFT approach} in the  $\mu/U_2$ and $1/U_2$ plane for $U_4/U_3=-3$ and $U_3/U_2=-2$. 
The MI lobes are denoted by the continuous lines and the dashed line separates the empty state. 
The SF to MI transitions are characterized by examining the 
behavior of change in the total density of the system $\rho$ and the superfluid density $\rho_s$ 
with respect to increase in chemical potential $\mu$. 
In the SF phase $\rho$ increases continuously with increase 
in $\mu$. However, in the MI phase $\rho$ remains constant for a range of $\mu$ and at the same time $\rho_s$ vanishes. In
Fig.~\ref{fig:rhomu1}(a) we show the $\rho - \frac{\mu}{U_2}$ plot {determined using the CMFT approach}
for various values of {$1/U_2=0.02,~0.04$ and $0.06$} which cut through different regions of the phase diagram of Fig.~\ref{fig:phasedia1} indicating the MI plateaus and the SF regions. 
The end points of the plateaus correspond to two different chemical potentials $\mu^+$ and $\mu^-$ of the system defined as 
\begin{equation}
 \mu^+=E_L(N+1)-E_L(N);~~ \\
 \mu^-=E_L(N)-E_L(N-1).
\end{equation}
Here, $E(N)$ denotes the ground state energy of the system with $N$ particles. The difference $G=\mu^+-\mu^-$ quantifies 
the gap in the MI phase which vanishes in the SF phase. The signatures of the 
MI and SF phases are also confirmed from the $\frac{\mu}{U_2}-\rho_s$ plot in Fig.~\ref{fig:rhomu1}(b) which shows finite(zero) 
superfluid density in the SF(MI) phase. 

Interesting thing happens in the regime of large interactions. It can be seen from Fig.~\ref{fig:corr1}
that for large $U_2=50${($1/U_2 = 0.02$), there are discrete jumps $\Delta\rho = 0.5$} in $\rho$(blue down triangles) with respect 
to increase in $\frac{\mu}{U_2}$ in the region between two plateaus corresponding to the MI(1) and MI(3) phases. This indicates a 
change in the particle number $\Delta N=2${, since we have $L=4$ in our CMFT calculation,} in the region which is a signature of pair formation. We can identify this 
phase as the PSF phase which can be confirmed from the pair correlation functions~\cite{singh1,Singhpsf,singh2}. To this end we compute 
the $n$-particle {nearest neighbor correlation functions using the CMFT approach defined as 
\begin{equation}
 \Gamma_{n} = \langle (a_i^{\dag})^n (a_j)^n\rangle.
 \label{eq:gamma}
\end{equation}
where $i$ and $j$ are the nearest neighbor site index of our four site cluster.}

In Fig.~\ref{fig:corr1} we also plot the correlation functions $\Gamma_{n}$ for $n=1,~2$ and $3$ corresponding to the 
single-(black circles), two-(red squares) and three-particles(green up triangles) respectively for different values of $\mu/U_2$ at a fixed $1/U_2=0.02$ of the phase diagram 
given in Fig.~\ref{fig:phasedia1}. 
It can be seen that at the 
plateau regions corresponding to the MI(1) and MI(3) phases, all the correlation functions are vanishingly small. 
However, for the values of $\rho$ away from the plateau regions i.e. $1 < \rho < 3$,  
$\Gamma_2$ clearly dominates over $\Gamma_1$ and $\Gamma_3$. This is a clear indication of the 
existence of the PSF phase which is sandwiched between the MI(1) and MI(3) lobes 
in the large $U_2$ regime as shown in Fig.~\ref{fig:phasedia1}. There exists a SF-PSF phase transition at these densities indicated by the green circles. 
\begin{figure}[t]
\begin{center}
\includegraphics[width=1\columnwidth]{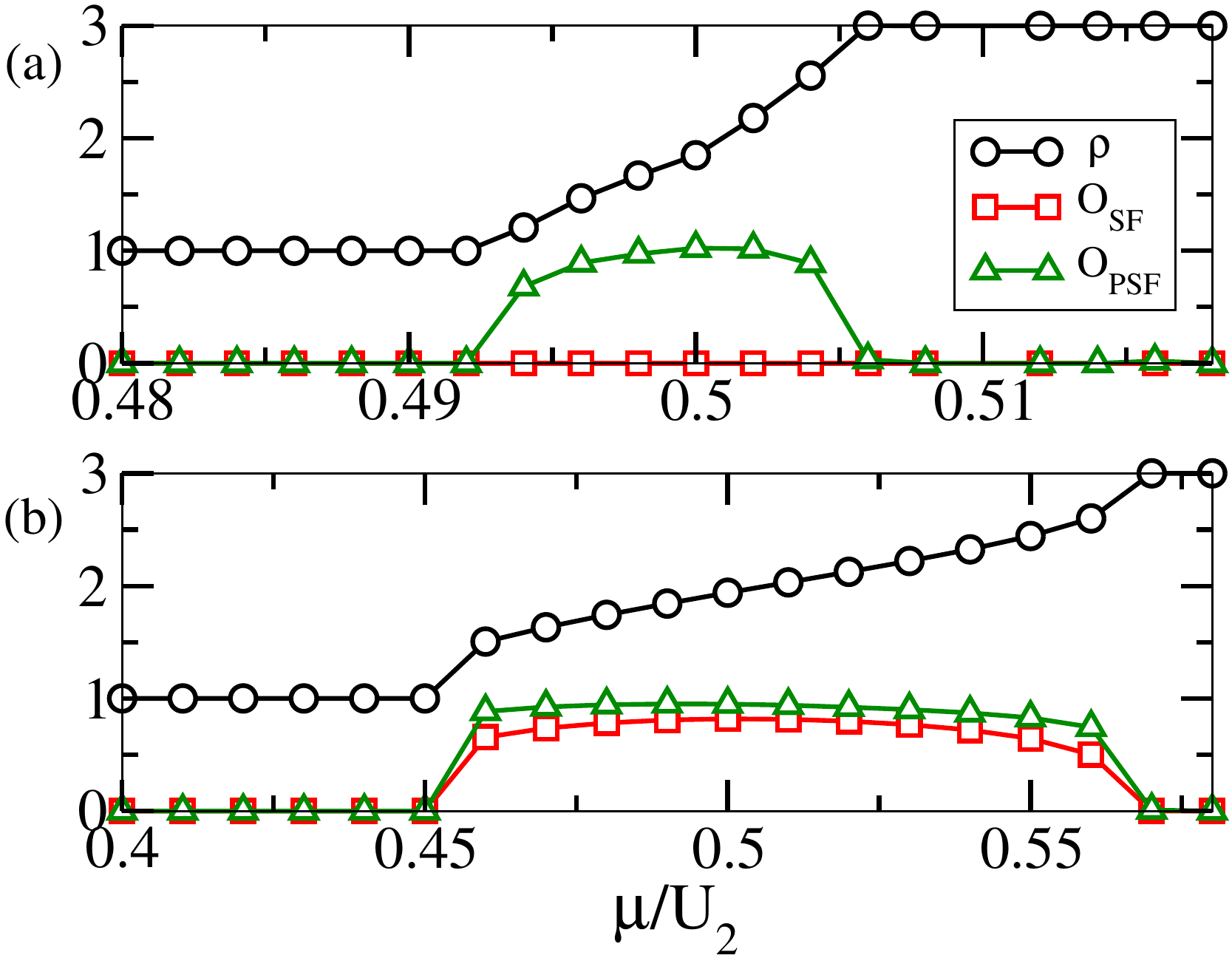}
 \end{center}
\caption{(color online)(a) iPEPS data for $\rho$, $O_{SF}$ and $O_{PSF}$ for $1/U_2=0.015$ showing the MI(1), MI(3) and the PSF phases. 
(b) Similar calculations for a cut passing through the normal Superfluid (SF) region at $1/U_2 = 0.04$. This phase is characterized by a 
non-vanishing value of both $O_{SF}$ as well as $O_{PSF}$ while the PSF phase is characterized by a vanishing $O_{SF}$ and non-zero $O_{PSF}$.}
\label{fig:corripeps1}
\end{figure}

As mentioned before, the CMFT approach can predict the quantum phases qualitatively and efficiently. 
However, to concretely establish the existence of the 
PSF phase of these two-body repulsively bound pairs we use the iPEPS method discussed before. 
In our simulation we use 
various physical quantities 
to identify different quantum phases. The gapped MI phases are identified by looking at the behavior of the 
chemical potential $\mu$ with respect to the average density $\rho$. The SF and the PSF phases are characterized by their respective order parameters defined as;
\begin{equation}
 O_{SF}=|\langle a_i\rangle|^2 
\end{equation}
and 
\begin{equation} 
O_{PSF}=|\langle a_i^2\rangle|^2.
\end{equation}
We compute these parameters for several values of $U_2$ and find signatures of different phases and phase transitions 
similar to that obtained using the CMFT method. In Fig.~\ref{fig:corripeps1}(a), we plot $\rho$(black circles), $O_{SF}$(red squares) and $O_{PSF}$(green triangles)
against $\mu/U_2$ for fixed $1/U_2=0.015$, $U_3/U_2=-2$ and $U_4 = \infty$. Note that the choice of $U_4$ restricts the local
Hilbert space to a maximum of three particles per site
and simplifies our iPEPS calculation while retaining the underlying physics of the system. 
It can be clearly seen from  Fig.~\ref{fig:corripeps1}(a) that there exists two Mott plateaus at $\rho=1$ and $3$ corresponding to the gapped MI(1) and MI(3) phases. Inside these plateau regions 
both the superfluid order parameters vanish. However, in the region between the two MI phases, the value of $O_{SF}$ remains vanishingly small, where as $O_{PSF}$ becomes finite 
indicating the existence of the PSF phase. 
We have also performed the same calculation for a different cut that 
passes through the region of normal superfluid (SF) as shown in Fig. \ref{fig:corripeps1} (b) for $1/U_2=0.04$. 
We find that both the $O_{SF}$ as well as the $O_{PSF}$ are non-zero in this region which defines our SF phase.

It can be noted that the physics obtained using the CMFT approach and the iPEPS method are similar to the one reported in Ref.~\cite{Singhpsf}. The 
important difference is the choice of the interactions. We explicitly show that in the presence of two-body repulsion the bosons prefer to move 
in pairs due to the large three-body attraction and four-body repulsion. The physics of the pair formation and the PSF 
phase on top of the MI(1) phase  can be understood from the energy consideration as discussed in Ref.~\cite{Singhpsf}. Due to the large 
three-body attraction the system will tend to acquire two particles at a time to reach the energy minimum by forming a trimer. However, because of the presence of 
uniform two body repulsion from the MI(1) background the added particles tend to move in pairs without affecting the system energy. This 
leads to the PSF phase in the system. This is indeed an interesting
manifestation of the multi-body interactions in the Bose-Hubbard model. We would like to mention that this pair formation is not limited to the region between the MI(1) and MI(3) lobes. 
One can in principle create the PSF phase 
between higher Mott lobes such as between the MI(2) and MI(4) lobes. To achieve this one needs to consider a five-body interaction term by keeping terms up to 
$M=5$ in the model given in Eq.~\ref{eq:ham1}. Using the CMFT calculation we verify that the PSF in this case can be obtained for suitable choice of 
repulsive $U_2,~U_3,~U_5$ and attractive $U_4$ terms in Eq.~\ref{eq:ham1}. Because of the attractive nature of $U_4$ in this case, 
the value of $U_5$ has to be very strong and repulsive to  prevent the collapse.

\subsection{Finite temperature analysis}
\begin{figure}[t]
\begin{center}
\includegraphics[width=1\columnwidth]{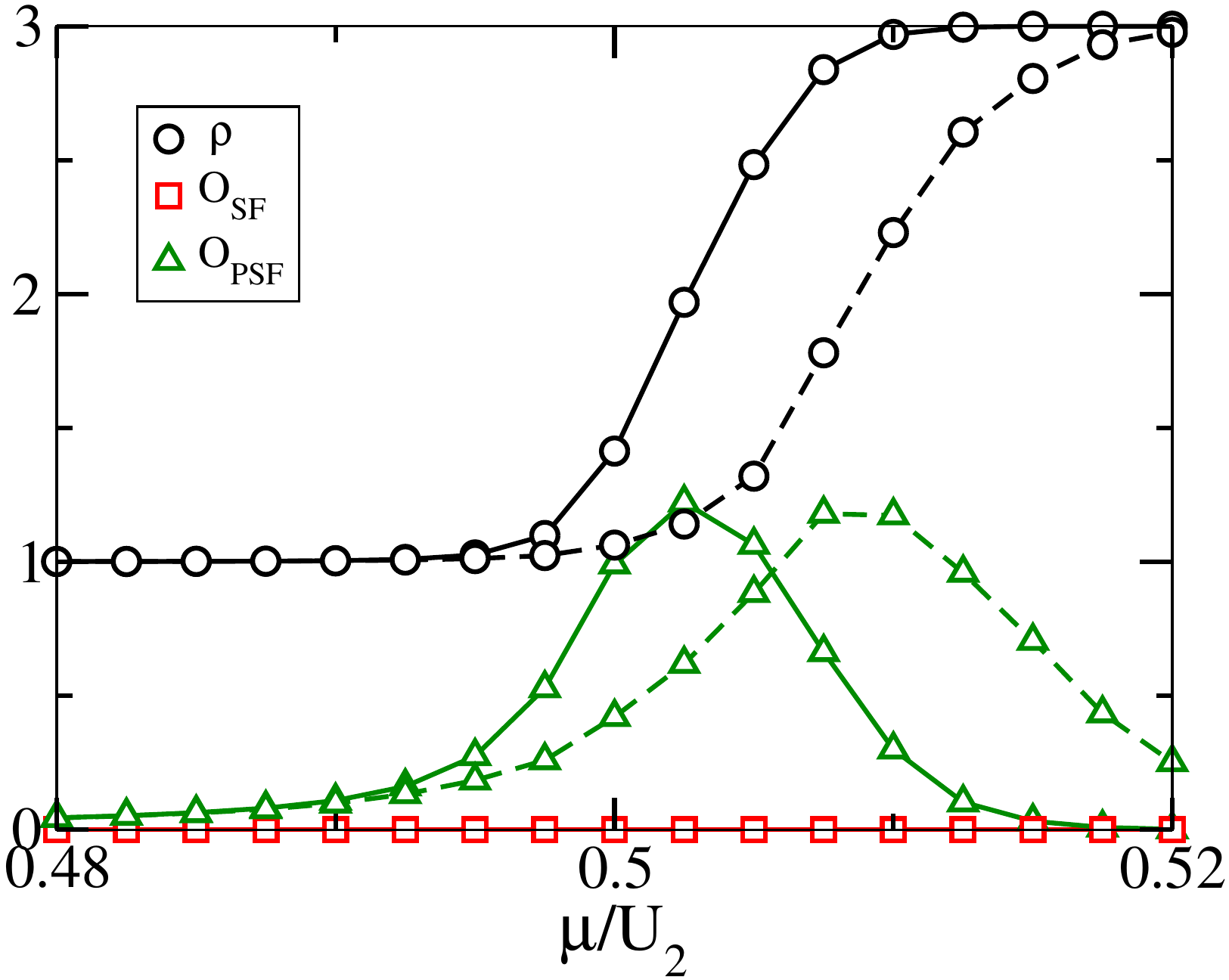}
\end{center}
\caption{(color online)iPEPS data for $\rho$, $O_{SF}$ and $O_{PSF}$ for $1/U_2=0.015$ at finite temperature. Here continuous and dashed lines 
are corresponding to temperature $T = 0.1$ and $0.2$ respectively where $T$ is the inverse of the thermodynamic $\beta$.}
\label{fig:corripeps2}
\end{figure}

After discussing the zero temperature phase diagram of the model shown in Eq.~\ref{eq:ham2} we embark on to analyze the effect of temperature on the 
system. As it is well known that the temperature is an unavoidable parameter in the real cold gas experiment~\cite{Gerbier2007,Kopec2014,wessel} and the quantum phases are fragile in presence 
of thermal fluctuation, it is pertinent to 
examine the stability of the PSF phase. At this stage, 
we perform finite temperature calculations using iPEPO to check the survival of the PSF phase by gradually increasing the system temperature $T$({$T$ is the inverse of the thermodynamic $\beta$}). 
We compute the different order parameters for these thermal states i.e. $O_{SF}$ and $O_{PSF}$ along with $\rho$ for the same choice of parameters considered in 
Fig.~\ref{fig:corripeps1}(a) for the zero temperature calculation and plot them in Fig. \ref{fig:corripeps2}. We show two different values of 
temperature such as $T=0.1$ and $T=0.2$ at which the PSF phase clearly survives which can be seen from the finite values of $O_{PSF}$. Above this 
temperature the PSF phase slowly disappear. This confirms that the PSF phase is stable against the thermal fluctuation. 

\subsection{Effect of density induced tunneling}
\begin{figure}[t]
\begin{center}
\includegraphics[width=1\columnwidth]{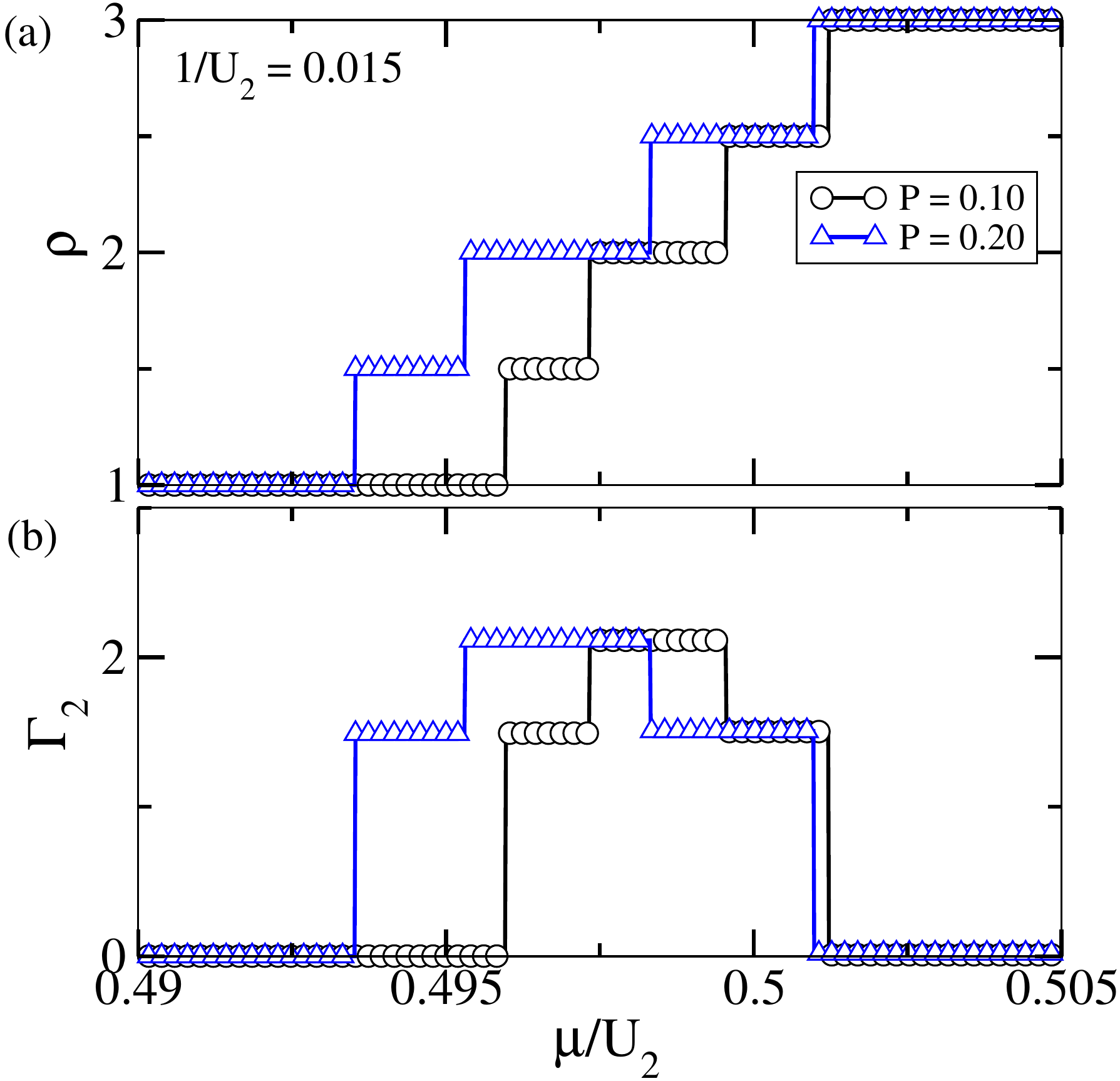}
\end{center}
\caption{(color online)The figure shows the existence of the PSF phase in the presence of density-induced tunneling($P$). Here we plot particle density($\rho$) in (a) and pair correlation function($\Gamma_2$) in (b) for $P=0.1$(black lines with circles) and $0.2$(blue lines with triangles). The other parameters are same as considered in Fig~\ref{fig:phasedia1} at $U_2=1/0.015$.}
\label{fig:ddt}
\end{figure}

In this subsection, we analyze the effect of density-induced tunneling on the PSF 
phase~\cite{OmjyotiLewenstein2015}. It has been theoretically shown that in optical 
lattice experiments, the density induced tunneling plays an important role and has been 
experimentally observed recently\tapan{~\cite{Jurgensen2014}}. Although the amplitudes of 
such tunneling are small compared to the conventional tunneling amplitude $t$ of the model 
shown in Eq.\ref{eq:ham2}, the natural question to ask is whether the narrow region of the 
PSF phase will survive in the presence of such density induced tunneling or not. In this 
context, we introduce the density-induced tunneling term in Eq.~\ref{eq:ham2} which is 
given by
\begin{equation}
 H_{P} = -P\sum_{\langle i,j \rangle}\left(a^\dagger_i(n_i+n_j)a_j +H.c.\right)
\end{equation}
where $P$ is the density-induced tunneling amplitude. Using the CMFT approach, we show that, indeed, the PSF phase survives up to a finite value of $P$. In the CMFT method, the density-dependent tunneling term can be decoupled as
\begin{equation}
 a^\dagger_i(n_i+n_j)a_j \approx \psi(a^\dagger_in_i + n_ja_j)
\end{equation}
where the terms O$(\psi^3)$ are neglected~\cite{Luhmann2012} and $\psi$ is the superfluid order parameter.

\begin{figure}[t]
   \centering
\includegraphics[width=1\columnwidth]{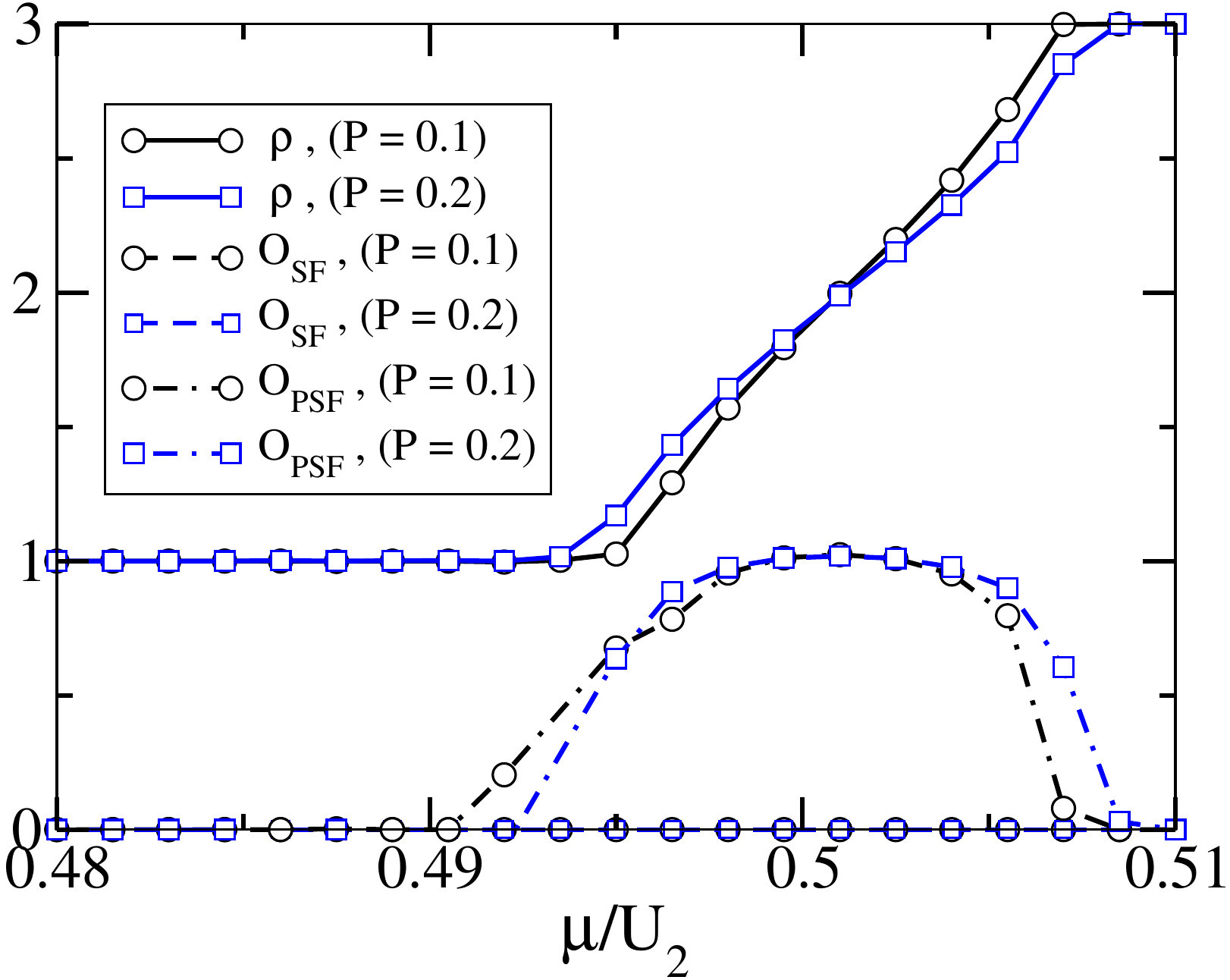}
\caption{(Color online) Figure shows the existence of PSF phase for different values of density dependent hopping amplitude $P$. Here we plot $\rho$(solid lines), $O_{SF}$(dashed lines) and $O_{PSF}$(dot-dashed lines) for $P=0.1$ and $0.2$, which indicate that the PSF survives for finite values of $P$.}
\label{fig:ddtipes}
\end{figure}

In Fig.~\ref{fig:ddt}, we plot the behavior of various physical quantities with respect to $P$ for a cut through the phase diagram of Fig.~\ref{fig:phasedia1} at $1/U = 0.015$ which passes through the PSF phase. It can be seen from Fig.~\ref{fig:ddt}(a), where we plot particle density($\rho$) for $P=0.1$ and $0.2$, that the $\rho$ exhibits discrete jumps in steps of $\delta \rho=0.5$, indicating a PSF phase. 
The corresponding pair correlation 
function($\Gamma_2$) are plotted in Fig.~\ref{fig:ddt}(b) which 
confirm the existence of the PSF phase for finite values of $P$. 
However, when we further increase the $P$, the system exhibite a normal SF phase for $P=0.3$.
We also analyze this 
situation using the iPEPS method which is 
shown in Fig.~\ref{fig:ddtipes}. The figure depicts that the $O_{PSF}$($O_{SF}$) is finite(zero) for finite values of $P$, indicating the existence of the PSF phase.

\section{Conclusions} 
In this paper we analyze a multi-body interacting Bose-Hubbard model and show the possibility of creating two-body repulsive bound bosonic pairs in 
a two dimensional optical lattice due to the combined effects of the multi-body interactions. 
We establish that for a very strong four-body repulsion a suitable ratio between the three-body attraction and two-body repulsion leads to the 
pair formation and hence the PSF phase between the MI(1) and MI(3) lobes. This fact is concretely demonstrated by analyzing the 
ground state properties of the BH model using the CMFT approach as well as the iPEPS method. Moreover, we show that this pair formation 
is stable against the thermal fluctuation and density induced tunneling effects which are inevitable in cold gas experiments. 
Due to the recent development in the field of ultracold quantum 
gas experiments, if it can be made possible to engineer the multi-body interactions among the bosons then it will be possible to create 
the repulsively bound pairs in an alternate way as opposed to the already observed one~\cite{Winkler2006}. Moreover, this finding may provide scope to 
create and manipulate the number of pairs in a controlled manner. 

\section{Acknowledgement}
TM acknowledges DST-SERB, India for the financial support 
through Project No. ECR/2017/001069. Part of the computational simulations were carried out 
using the computing facilities of Param-Ishan at Indian Institute of Technology - Guwahati, India.

\bibliography{reference}

\end{document}